\documentstyle[12pt,aaspp4,flushrt]{article}

\received{4 February 1997}
\accepted{9 April 1997}

\slugcomment{To appear in the Astrophysical Journal Letters}

\lefthead{Bower et al.}
\righthead{Nucleus of M84}

\begin{document}

\title{The Nuclear Ionized Gas in the
Radio Galaxy M84 (NGC~4374)\altaffilmark{1}}

\author{Gary A. Bower\altaffilmark{2}, Timothy M. Heckman\altaffilmark{3,4},
Andrew S. Wilson\altaffilmark{4,5}, and Douglas O. Richstone\altaffilmark{6}}






\vskip 1.0in

\centerline{received: 4 February 1997; accepted: 9 April 1997}


\altaffiltext{1}{Based on observations with the NASA/ESA {\it Hubble Space Telescope},
obtained
at the Space Telescope Science Institute, which is operated by the
Association of Universities for Research in Astronomy, Inc. (AURA), under
NASA contract NAS5-26555.}
\altaffiltext{2}{Kitt Peak National Observatory, National Optical Astronomy Observatories, P. O. Box 26732,
Tucson, AZ 85726; gbower@noao.edu}
\altaffiltext{3}{Department of Physics \& Astronomy, Johns Hopkins University,
Homewood Campus, Baltimore, MD 21218; heckman@pha.jhu.edu}
\altaffiltext{4}{Space Telescope Science Institute, 3700 San Martin Drive,
Baltimore, MD 21218}
\altaffiltext{5}{Astronomy Department, University of Maryland, College Park,
MD 20742; wilson@astro.umd.edu}
\altaffiltext{6}{Department of Astronomy, University of Michigan, Dennison Building, Ann
Arbor, MI 48109; dor@astro.lsa.umich.edu}


\newpage

\begin{abstract}

We present optical images of the nucleus of the nearby
radio galaxy M84 (NGC~4374 = 3C272.1) obtained with the
Wide Field/Planetary Camera 2 (WFPC2) aboard the {\it Hubble Space Telescope} (HST). Our three images cover the H$\alpha$ + [N~II] emission lines as
well as the $V$ and $I$ continuum bands. Analysis of these images confirms that the H$\alpha$ + [N~II]
emission in the central $5''$ (410 pc) is elongated along position angle
(P.A.) $\approx 72\arcdeg$, which is roughly parallel to two nuclear dust lanes.
Our high-resolution images reveal that
the H$\alpha$ + [N~II] emission has three components, namely a nuclear gas disk,
an
`ionization cone', and outer filaments.
The nuclear disk of ionized gas has
diameter $\approx 1'' = 82$ pc and major axis P.A.
$\approx 58\arcdeg \pm 6\arcdeg$. On an angular scale of $0\farcs5$, the major axis of this nuclear gas disk is consistent with
that of the dust. However, the minor axis
of the gas disk (P.A. $\approx 148\arcdeg$) is tilted with respect to that of the filamentary
H$\alpha$ + [N~II] emission at distances $> 2''$ from the nucleus; the minor
axis of this larger scale gas
is roughly aligned with the axis of the kpc-scale radio jets (P.A. $\approx
170\arcdeg$).
The ionization cone (whose apex is offset by $\approx 0\farcs3$ south of the nucleus)
extends $2''$ from the nucleus along the axis of the southern radio jet. This feature is similar to the ionization
cones seen in some Seyfert nuclei, which are also aligned with the radio axes.    

\end{abstract}


\keywords{galaxies: active --- galaxies: elliptical --- 
galaxies: individual (M84) --- galaxies: jets ---
galaxies: nuclei}


%

\newpage   

\section{Introduction}

Recent high-resolution imaging surveys 
of nearby
elliptical galaxies with HST (e.g., Jaffe et al. 1994; Lauer et al.~1995) have found that
small ($\lesssim 200$ pc) axisymmetric structures in the nuclear regions are common. These structures
include filaments or disks
of dust and ionized gas. The identification
of such disks in galaxies hosting an active galactic
nucleus (AGN) is especially interesting, since AGNs are thought to be
powered by the accretion of gas into the relativistically deep potential
well of a supermassive black hole. In the context of this model, powering an
AGN for a long duration requires that large amounts of gas in the galaxy's ISM be transported
to the nucleus for eventual accretion onto the central black hole. It is unlikely
that this gas would arrive at the nucleus with no angular momentum, so the
nuclear gas is expected to form a thin disk, which is
thought to be responsible for collimating the jets in 
AGNs in a direction approximately perpendicular to the plane of the disk (Rees 1984).

Discovery of ionized gas disks in nearby radio galaxies is especially 
important because the kinematics of such disks might provide an estimate of the nuclear mass, if the gas exhibits Keplerian motion about
the nucleus. Since high spatial resolution ($\lesssim 10$ pc) is required for
such an analysis, the nearest radio galaxies, including the Virgo
cluster elliptical galaxies M87 and M84, are the best targets. 
The nucleus of M87 is known to contain an ionized
gas disk (Ford et al.~1994), the kinematics of which suggest the presence
of a $2 \times 10^9 \ M_{\odot}$ black hole, if the
gas is in circular motion (Harms et al.~1994).
For M84, previous narrow-band imaging obtained by
Hansen et al.~(1985) and Baum et al.~(1988) has shown that the
H$\alpha$ + [N~II] $\lambda\lambda 6548, 6583$ emission in the circumnuclear
region is $\approx
7'' \times 20''$ (570 pc $\times$ 1640 pc) in extent, and 
elongated in P.A. $\approx 83\arcdeg$.  This elongated emission region is roughly perpendicular to the radio jet axis, which is at P.A. $\approx 10\arcdeg$ on the pc scale (Jones et al.~1981) and
$\approx 170\arcdeg$ on the kpc scale (Laing \& Bridle
1987). However, the morphology of the ionized gas is not clearly resolved by ground-based
imaging. It could be a gas disk or filamentary emission associated with a
cooling flow (Hansen et al.~1985). The kinematics of the
H$\alpha$ + [N~II] emission region indicate that the gas is rotating about the
nucleus (Baum et al.~1990, 1992) and that the rotation gradient across the
nucleus is spatially unresolved (i.e., $>$ 100 km~s$^{-1}$ arcsec$^{-1}$), suggesting
the presence of a high mass concentration at the nucleus. These data provide a tantalizing clue
that M84 might indeed contain a nuclear gas disk. We have obtained higher resolution images of M84 with HST to determine if this preliminary inference is
correct. 

Throughout this paper, we adopt a distance to M84 of 17 Mpc 
(Mould et al.~1995). At this distance, 
$1''$ corresponds to 82 pc. The 
Galactic extinction along the line of sight is
$A_B = 0.13$ mag
(Burstein \& Heiles 1984).

\section{Observations and Data Reduction}

Images of the nuclear region of M84 were obtained with WFPC2
(Burrows 1995)
aboard HST on 1996 March 4 with the telescope tracking in fine lock 
(nominal \ jitter $\approx 0\farcs007$).
The nucleus was placed in the Planetary
Camera, which has a scale of $0\farcs044$ pixel$^{-1}$ and provides a
resolution of $\approx 0\farcs1$ (8 pc). Two exposures were obtained with each
of the filters F547M, F814W, and F658N, whose effective wavelengths/bandpass
widths are 5454~\AA/486~\AA, 8269~\AA/1758~\AA, and
6590~\AA/28.5~\AA. These filters provide images in H$\alpha$ $\lambda 6563$ +
 [N~II]
$\lambda\lambda 6548, 6583$ and the neighboring continua. Total exposure
times through these filters were 1200 sec, 520 sec, and 2600 sec,
respectively. Initial data reduction was accomplished by the HST
pipeline software, after which the two sub-exposures in each filter were
combined to reduce the effect of cosmic ray events.  
We adopted the flux calibration provided by the HST pipeline software.
For the F547M and F814W images, we have transformed the instrumental magnitudes
to Landolt V and I magnitudes, respectively, by using the calcphot task in the synphot package in
STSDAS on a synthetic early-type galaxy spectrum from Bica (1988) that closely
matches the spectrum of the stellar population in M84. Although the
magnitudes that we quote in the present paper are in the Landolt system, the fluxes quoted are
at the effective wavelengths given above.
To register the images, we measured the positions 
of $\approx 20$ globular clusters
in M84 in each of the images and found no
rotational shifts and only small translational
shifts ($\approx 0.5-1.0$ pixel) between the three images. After correcting
for these shifts, the images are aligned to $\pm 0.1$ pixel ($\pm 0\farcs004$). 
To the accuracy of alignment, the bright central point source is coincident in all three bandpasses. Thus,
we adopt the central continuum peak as the location of the nucleus. Since this galaxy contains two prominent
dust lanes near the nucleus (see \S 3), removing the continuum contribution from
the F658N image to obtain the H$\alpha$ + [N~II] image is more difficult
than simply subtracting the F547M or F814W image. We constructed a
synthetic image of the continuum at 6590~\AA \ using the F547M image and the
ratio of the F547M and F814W images in conjunction with the assumptions that
(1) the unreddened spectrum of the stellar population matches that of the
synthetic early-type galaxy spectrum used above, and (2) the internal reddening
associated with the prominent dust lanes is described by the Galactic 
interstellar extinction curve. This synthetic 6590~\AA \ continuum image
was subtracted from the F658N image, yielding the H$\alpha$ + [N~II] image.
Since the individual exposure times
of the two sub-exposures obtained in the filter F658N were long (1000 sec
and 1600 sec), there were $\approx 100$ pixels that were affected by
cosmic ray events in both sub-exposures and thus could not be removed
by combining them. We removed these residual cosmic ray events from the
H$\alpha$ + [N~II] image by 
applying a median filter with dimensions of $3 \times 3$ pixels. The resolution
of this smoothed image was determined to be $\approx 0\farcs13$ by comparing
the FWHM of a model PSF (Krist~1995) before and after it had been smoothed
with an identical median filter. This median filter conserved flux in the
image to $\lesssim 1$\%, except for pixels within $0\farcs13$ of
the nucleus, where the flux loss was typically 30\%.
We converted the flux densities $F_{\lambda}$ in the H$\alpha$ + [N~II] image
to fluxes integrated over these emission lines by using the synphot task
calcphot. We calculated values of $F_{\lambda}$ for the F658N bandpass both for the synthetic 
early-type galaxy spectrum (see above) alone and for this synthetic spectrum
with model H$\alpha$ and [N~II] $\lambda\lambda 6548, 6583$ emission lines
(whose relative line ratios were taken from Hansen et al.~1985) added.
The difference in these predicted values of $F_{\lambda}$ was compared with
the H$\alpha$ + [N~II] flux in the model emission lines to show that
the values of $F_{\lambda}$ in the H$\alpha$ + [N~II] image should be multiplied by 25.7 (which is close to the
effective bandwidth of the F658N filter of 28.5 \AA) to obtain total
H$\alpha$ + [N~II] fluxes.

\section{Results}

Fig.~1 (Plate X) shows
grayscale representations of the F547M, F814W, (V$-$I), and H$\alpha$ + [N~II] images. The continuum images show a compact source at
the nucleus (which lies at the photometric center of the galaxy), as well as two dust lanes 
(also seen by Jaffe et al.~1994) within the
central $5''$ whose orientation is roughly parallel to the filamentary 
H$\alpha$ + [N~II]
emission extended on the same scale. Both dust lanes are oriented roughly in an east-west direction at $\ge 1''$ east of
the nucleus. However, west of this point, they both bend toward the southwest 
by $\approx 25\arcdeg$. 

Fig.~1(f) shows that the H$\alpha$ + [N~II]
emission closest to the nucleus has a very interesting
distribution. A contour plot of this image (Fig.~2) shows that 
the isophote shapes change with increasing distance $R$ from the nucleus.
For $R < 0\farcs5$ (41 pc), the isophotes are nearly elliptical, as would be
expected for an inclined, circular thin gas disk. 
Baum et al.~(1990)
found that the gradient of the rotational velocity across the nucleus is spatially unresolved, with the projected rotation axis lying at
P.A. $0\arcdeg$ (Baum et al.~1990) or P.A. $24\arcdeg$ (Baum et al.~1992).
This rotation axis direction is broadly consistent with the photometric
minor axis of the gas distribution for $R > 2''$, suggesting that the
gas forms a rotating disk.
To determine the geometrical properties of this disk,
we fitted model ellipses to
these elliptical isophotes using the `ellipse' task (see Jedrzejewski 1987)
in the STSDAS package. During the fit, we allowed the ellipse center, intensity,
position angle, and ellipticity to be free parameters. Successful fits were
obtained for $R < 0\farcs5$, but outside this radius ellipses cannot fit
the isophotes since they have 
irregular shapes. 
The main results of the fits to these inner isophotes are:

\noindent
(1) The isophotes are concentric to $\pm 0\farcs05$ (4 pc), and the
center of the gas disk coincides with the nucleus (and with the
H$\alpha$ + [N~II] peak; see \S 2). This alignment in M84 contrasts with the 
offset of $13 \pm 7$ pc between the centers of the galaxy and dust disk in the radio
galaxy NGC~4261 (Ferrarese et al.~1996).

\noindent
(2) The average P.A. and ellipticity for radii $0\farcs25 - 0\farcs5$ are $58\arcdeg \pm 6\arcdeg$ and $0.17 \pm 0.07$, respectively
(the quoted errors are dominated by the noise in the image). The P.A. and ellipticity
measurements for $R < 0\farcs25$ were not included in these averages because
the finite resolution and pixel sampling make the isophotes circular.
The minor axis of the disk is not quite parallel to the radio jet axis,
but is offset by $22\arcdeg$. By comparison, the minor axis of the gas disk in
M87 is offset by $19\arcdeg$ from its 
radio jet axis (Ford et al.~1994). If the disk is thin and circular, the
above ellipticity implies an inclination of $34\arcdeg^{+7}_{-8}$.

\noindent
(3) The H$\alpha$ + [N~II] emission at the nucleus is very compact.
Fig.~3 shows a comparison of the intensity profile given by the isophote
fits with the shape of a 
model PSF generated by the program Tiny Tim (Krist~1995). A  
$3 \times
3$ median filter has also been applied to this model PSF, similar to the H$\alpha$ + [N~II] image
(see \S 2).
The total flux of the normalized PSF is $8.7 \times 10^{-15}$
erg cm$^{-2}$ s$^{-1}$.

The observed spatial distribution of the dust closest to the nucleus indicates that it lies in the nuclear region (rather than simply lying in the foreground). A comparison of
Figs.~1(a) and 1(c) shows that the compact nuclear continuum source (discussed in more detail below) is bluer than the other reddened
regions in the central $1''$. Fig.~1(e) shows the (V$-$I) color map of a $3'' \times
3''$ region centered on the nucleus. In this region, the reddening
is greatest along the northern edge of the central dust lane and at the patch of
dust centered on the nucleus
with diameter $\approx
0\farcs3$. The nuclear dust lane lies along the same P.A. as the gas
disk, suggesting that it is associated with the gas disk.
This dust morphology is also seen in the
F547M and F814W images when displayed with suitable intensity scales,
although not as prominently as in Fig.~1(e) because of the
bright nuclear continuum source.      

The H$\alpha$ + [N~II] emission outside the central gas disk appears to have
two components (see Figs.~1(d) and 1(f)). The most prominent component is the filamentary emission
extended parallel to the dust lanes, while the fainter component is a
possible conical structure extending $\approx 2''$ south from the nucleus 
with axis along 
P.A. $\approx 162\arcdeg$, which is consistent with the axis of the
kpc-scale radio jet
(P.A. $\approx 170\arcdeg$). This cone has a well defined, sharp, straight
edge on at least the NE edge, suggesting shadowing of a nuclear source.
There is also a hint of a similar conical structure to the north of the nucleus. Although the geometrical
properties of the south cone are somewhat subjective, 
we find an opening angle of $\approx 73\arcdeg$ at the apex (which is
located $\approx 0\farcs3$ south of the nucleus). 
The general appearance of these conical structures in
our H$\alpha$ + [N~II] image of M84 resembles the ionization cones seen in some Seyfert galaxies (Wilson \& Tsvetanov
1994 and references therein). The axis of M84's conical structures is well-aligned
with the radio jet axis, similar to the
strong alignment seen in Seyfert galaxies between these structures.
The ionization cone in M84 is more closely aligned with the kpc-scale
radio jets than with the axis of the pc-scale jet (P.A. $\approx 190\arcdeg$). Perhaps
this suggests that the structure shadowing the nuclear source is located on
scales $>> 1$ pc. However, confirmation of this ionization cone
requires HST imaging of M84 in a high-excitation emission line (e.g., [O~III] 
$\lambda 5007$).    

The apparent flux and color of the nuclear continuum source seen in Fig.~1 (also seen in an archival WFPC image; Jaffe et al.~1994) can
be measured by fitting a model PSF 
(constructed with Tiny Tim; Krist 1995) to the nucleus.
We determined
the normalization of the model PSF to be such that subtracting it from the
original image would yield an image in which the intensity is roughly
constant within $\approx 0\farcs3$ of the nucleus.
The flux and color (both corrected for Galactic extinction but
not internal extinction) of the nuclear continuum source implied by this
procedure are V = 19.9 and (V$-$I) = 1.6. These values are clearly
affected by extinction and reddening by dust within M84.

M84's dust and gas content can be estimated from our (V$-$I) color map, assuming
that light from the stars lying in the foreground of the embedded dust lanes
is insignificant and that the
extinction internal to M84 follows the Galactic interstellar extinction
curve and gas-to-dust ratio. Fig.~4 shows a histogram of the (V$-$I) values 
from
Fig.~1(c) (after correcting for Galactic reddening) for only the region of the image encompassing the dust lanes.
This region is represented by a rectangle $14\farcs5$ wide in right ascension
and $8\farcs4$ high in declination (whose center lies $2\farcs2$ NNE of the
nucleus). Although the (V$-$I) at the peak of the distribution
in Fig.~4 is 1.4, this probably is not representative of the unreddened color.
Buta \& Williams~(1995) found that (V$-$I) $= 1.24 \pm 0.008$ inside the
half-light radius of M84,
while the mean (V$-$I) for elliptical galaxies with
the same Hubble type as M84 is 1.2. Adopting the latter value as the unreddened
color of M84, the mean 
(V$-$I) of 1.43 indicates a mean internal extinction of $\langle A_V \rangle = 0.54$ within the region encompassing the dust lanes.
This provides an estimate of the mass in dust using $M_d = \Sigma \langle A_V \rangle/ \Gamma_V$
(e.g., van Dokkum \& Franx~1995; Sadler \& Gerhard~1985), where $\Sigma$
is the surface area affected by dust extinction and $\Gamma_V$ is the
visual mass absorption coefficient $\approx 6 \times 10^{-6}$ mag kpc$^2$
$M_{\odot}^{-1}$. The area of the rectangle is $\Sigma = 0.82$ kpc$^2$, so
$M_d = 7 \times 10^4
\ M_{\odot}$. This agrees with the dust mass determined from the far-infrared fluxes measured by IRAS (e.g., Roberts et al.~1991; adjusted to
our adopted distance), suggesting that our color map of M84 provides an
adequate measure of the dust content. This agreement between optical and 
far-infrared dust masses contrasts with the optical dust ``deficit'' often
found in other optical studies of elliptical galaxies when dust whose spatial distribution closely follows that of
the stars is missed
(Goudfrooij \& de Jong~1995). Adopting the Galactic gas-to-dust
ratio ($M_{\rm gas}/M_d \approx 130$), the total mass of gas and dust within
this central region of M84 is $9 \times 10^6 \ M_{\odot}$.    
The mass in ionized gas
is a very small fraction of this total. The total flux in H$\alpha$ (corrected for Galactic and internal extinction)
is
$6.3 \times 10^{-14}$ erg cm$^{-2}$ s$^{-1}$ if we adopt the H$\alpha$/[N~II]
$\lambda\lambda 6548, 6583$ ratio measured by Hansen et al.~(1985). Assuming case B recombination,
$T_e = 10^4$ K, and $N_e = 10^3$ cm$^{-3}$, we find $M(H^+) \approx
6 \times 10^3 \ M_{\odot}$. Most of the gas must thus be neutral or molecular.
Although searches for emission from H I 21 cm and
CO ($2-1$) 1.3 mm have been unsuccessful (Huchtmeier 1994; Knapp \& Rupen 1996),
the upper limits (e.g., $M(H I) < 5 \times 10^8 \ M_{\odot}$) are not
significant.

\acknowledgments

We acknowledge useful discussions with Richard Green, Vicki Sarajedini,
Luis Ho, and an anonymous referee.
Support for this work was provided by NASA through grant number GO-6094 to
the Johns Hopkins University from the Space Telescope Science Institute,
which is operated by AURA, Inc., under NASA contract NAS5-26555, and
grants NAGW-3268 and NAGW-4700.

\clearpage

%
%

\par\noindent
{\bf References}

\par\noindent
Baum, S. A., et al.~1988, \apjs, 68, 643
\par\noindent
Baum, S. A., Heckman, T., \& van Breugel, W.
1990, \apjs, 74, 389
\par\noindent
Baum, S. A., Heckman, T., \& van Breugel, W.
1992, \apj, 389, 208
\par\noindent
Bica, E. 1988, \aap, 195, 76
\par\noindent
Burrows, C. J. (ed) 1995, Wide Field and Planetary Camera 2 Instrument

Handbook V3.0,
(Space Telescope Science Institute, Baltimore, MD)
\par\noindent
Burstein, D., \& Heiles, C. 1984,
\apjs, 54, 33
\par\noindent
Buta, R., \& Williams, K. L.~1995, \aj, 109, 543
\par\noindent
Ferrarese, L., Ford, H. C., \& Jaffe, W.~1996, \apj, 470, 444 
\par\noindent
Ford, H. C., et al.~1994, \apj, 435, L27
\par\noindent
Goudfrooij, P., \& de Jong, T.~1995, \aap, 298, 784
\par\noindent
Hansen, L., N\o rgaard-Nielsen, H. U., \& 
J\o rgensen, H. E. 1985, \aap, 149, 442
\par\noindent
Harms, R. J., et al.~1994, \apj, 435, L35
\par\noindent
Huchtmeier, W. K.~1994, \aap, 286, 389 
\par\noindent
Jaffe, W., et al.~1994, \aj, 108, 1567
\par\noindent
Jedrzejewski, R. I. 1987, \mnras, 226, 747
\par\noindent
Jones, D. L., Sramek, R. A., \& Terzian, Y.~1981, ApJ, 246, 28
\par\noindent
Knapp, G. R., \& Rupen, M. P.~1996, \apj, 460, 271
\par\noindent
Krist, J. 1995, in Astronomical Data Analysis Software
and Systems IV,
ed. 

R. A. Shaw, H. E. Payne, \& J. J. E. Hayes (Astronomical Society
of the Pacific, San 

Francisco, CA), 349
\par\noindent
Laing, R. A., \& Bridle, A. H. 1987,
\mnras, 228, 557
\par\noindent
Lauer, T. R., et al.~1995, \aj, 110, 2622
\par\noindent
Mould, J., et al.~1995, ApJ, 449, 413
\par\noindent
Rees, M. J.~1984, \araa, 22, 471
\par\noindent
Roberts, M. S., Hogg, D. E., Bregman, J. N., Forman, W. R., \& Jones,
C.~1991, \apjs, 75, 

751
\par\noindent
Sadler, E. M., \& Gerhard, O. E.~1985, \mnras, 214, 177
\par\noindent
van Dokkum, P. G., \& Franx, M.~1995, \aj, 110, 2027
\par\noindent
Wilson, A. S., \& Tsvetanov, Z. I.~1994, \aj, 107, 1227

%
%

\clearpage

\figcaption{HST/WFPC2 images of the nuclear region of M84, with resolution $\approx 0\farcs1$ (8 pc), except for the
H$\alpha$ + [N~II] image in (d) and (f), which has resolution $\approx 0\farcs13$. All images have the same orientation (north 
up and east to the left). Panels (a) $-$ (d) have the same scale, while (e) and (f) are expanded views of the nucleus. (a) The F547M (5454~\AA) image, and (b) the F814W (8269~\AA)
image.
The displayed intensities in both (a) and (b) (in units of erg cm$^{-2}$ s$^{-1}$ \AA$^{-1}$
arcsec$^{-2}$) 
range logarithmically from $1.2 \times 10^{-16}$ to $1.9 \times 10^{-15}$,
while the peak intensities are $3.1 \times 10^{-15}$ and $3.6 \times 10^{-15}$, respectively. (c) The (V$-$I) color map, obtained from the ratio of (a) to (b). The 
displayed color values range linearly from 
$F_{\lambda}(5454$~\AA)/$F_{\lambda}(8269$~\AA) of 0.65 to 1.1, which corresponds to
a range in (V$-$I) of 1.8 to 1.2. Darker shades represent redder colors. 
(d) The H$\alpha$ +
[N~II] image. The displayed intensities (in units of erg cm$^{-2}$ s$^{-1}$
arcsec$^{-2}$) range logarithmically from $7.9 \times 10^{-16}$ to
$2.6 \times 10^{-14}$, while the peak intensity is $1.9 \times 10^{-13}$.
(e) The central $3'' \times 3''$ of the (V$-$I) color
map presented in (c). The color values displayed range linearly from
$F_{\lambda}(5454$~\AA)/$F_{\lambda}(8269$~\AA) of 0.6 to 1.1 (i.e.,
(V$-$I) of 1.9 to 1.2), with darker shades representing redder colors (as in
(c)). (f) The central $5'' \times 5''$ of the
H$\alpha$ + [N~II] image presented in (d). The displayed intensities (in
units of erg cm$^{-2}$ s$^{-1}$
arcsec$^{-2}$) range logarithmically from $7.9 \times 10^{-16}$ to
$3.9 \times 10^{-14}$. 
 \label{fig1}}

\figcaption{A contour plot of the H$\alpha$ + [N~II] image in
Fig.~1(f), with the contour levels (in units of erg cm$^{-2}$ s$^{-1}$ arcsec$^{-2}$)
ranging from $3.0 \times 10^{-15}$ to $1.9 \times 10^{-13}$ (i.e., the
peak intensity) with an
interval of 0.5 mag. North is up and east to the left. \label{fig2}}

\figcaption{The intensity of H$\alpha$ + [N~II] in erg cm$^{-2}$ s$^{-1}$ arcsec$^{-2}$
as a function of distance $R$ from the nucleus along the disk's major axis
(open circles). 
For comparison, the points connected by a line show a model PSF normalized so that its peak intensity matches the peak intensity
of the observed H$\alpha$ + [N~II] emission. \label{fig3}}

\figcaption{A histogram showing the number of PC2 pixels (N$_{\rm pix}$)
with a given value of (V$-$I) in Fig.~1(c) (after correcting for Galactic reddening). This histogram includes only the region of the image encompassing the dust lanes (see text). \label{fig4}}   

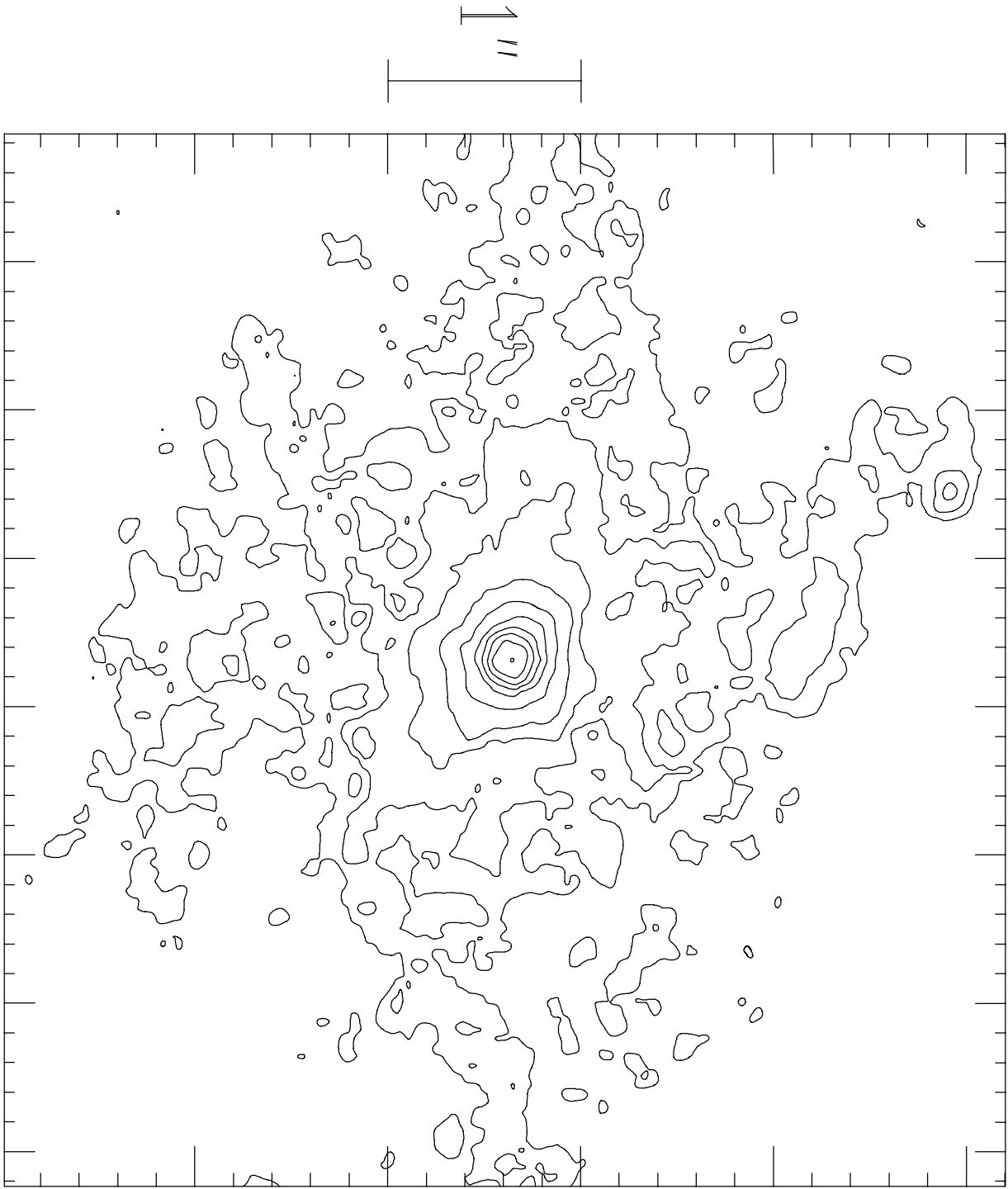
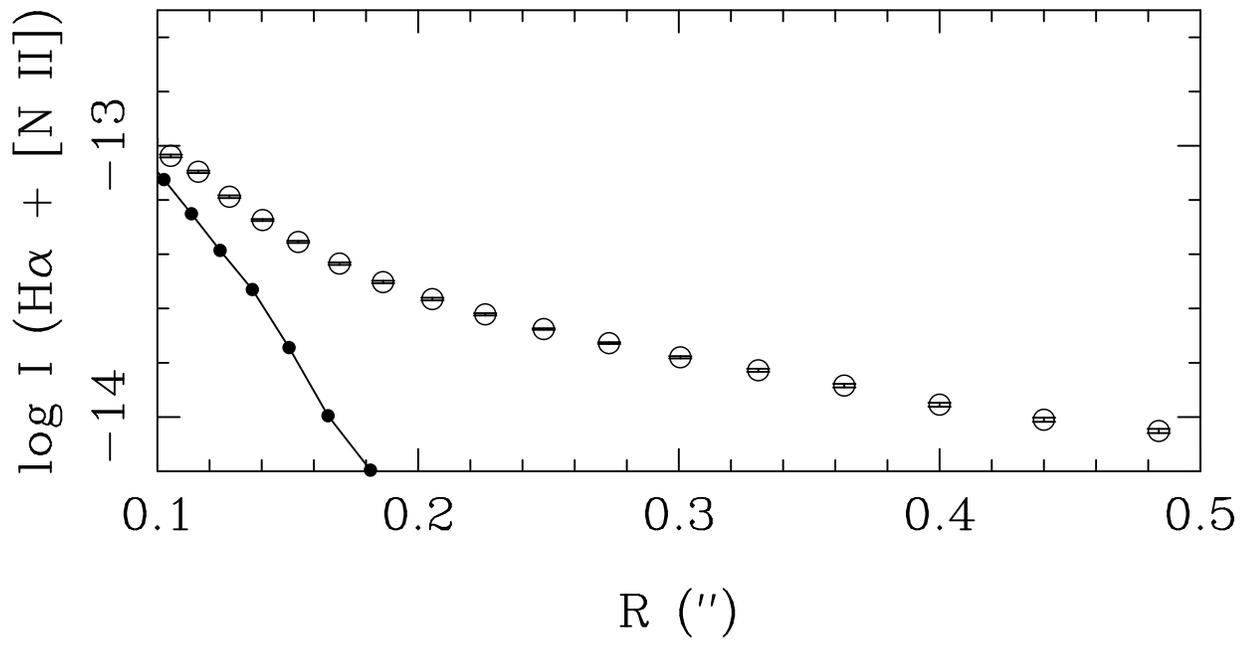
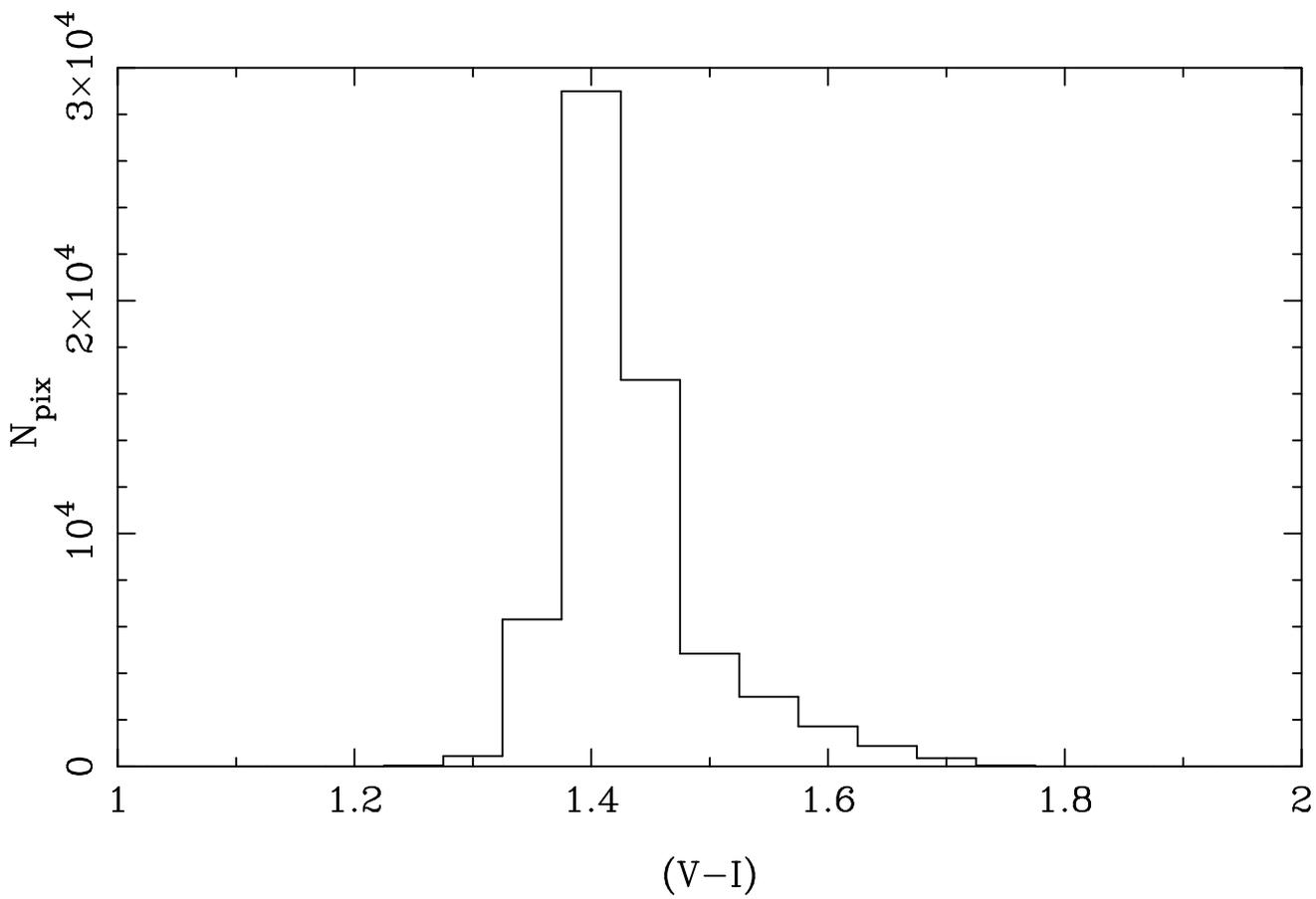
\end{document}